\documentclass[aps,prd,onecolumn,groupedaddress,showpacs,nofootinbib,amssymb
]{revtex4}
\usepackage[dvips]{graphicx,color}
\usepackage{amssymb}
\usepackage{amsmath}
\usepackage{graphicx}
\usepackage{amsfonts}
\usepackage{bm}

\newcommand{\be}{\begin{equation}}
\newcommand{\ee}{\end{equation}}
\newcommand{\bea}{\begin{eqnarray}}
\newcommand{\eea}{\end{eqnarray}}
\newcommand{\beaa}{\begin{eqnarray*}}
\newcommand{\eeaa}{\end{eqnarray*}}

\newcommand{\nn}{\nonumber \\}
\newcommand{\e}{\mathrm{e}}

\begin{document}

\title{Unimodular-Mimetic Cosmology}
\author{S.~Nojiri,$^{1,2}$\,\thanks{nojiri@gravity.phys.nagoya-u.ac.jp}
S.~D.~Odintsov,$^{3,4}$\,\thanks{odintsov@ieec.uab.es}
V.~K.~Oikonomou,$^{5,6}$\,\thanks{v.k.oikonomou1979@gmail.com}}
\affiliation{$^{1)}$ Department of Physics, Nagoya University, Nagoya 464-8602,
Japan \\
$^{2)}$ Kobayashi-Maskawa Institute for the Origin of Particles and the
Universe, Nagoya University, Nagoya 464-8602, Japan \\
$^{3)}$Institut de Ciencies de lEspai (IEEC-CSIC),
Campus UAB, Carrer de Can Magrans, s/n\\
08193 Cerdanyola del Valles, Barcelona, Spain\\
$^{4)}$ ICREA, Passeig LluAs Companys, 23,
08010 Barcelona, Spain\\
$^{5)}$ Tomsk State Pedagogical University, 634061 Tomsk, Russia\\
$^{6)}$ Laboratory for Theoretical Cosmology, Tomsk State University of Control Systems
and Radioelectronics (TUSUR), 634050 Tomsk, Russia
}

\tolerance=5000

\begin{abstract}
We combine the unimodular gravity and mimetic gravity theories into a unified theoretical framework, which is proposed to provide a suggestive proposal for a framework that may assist in the discussion and solution search of the cosmological constant problem and of the dark matter issue. After providing the formulation of the unimodular mimetic gravity and investigating all the new features that the vacuum unimodular gravity implies, by using the underlying reconstruction method, we realize some well known cosmological evolutions, with some of these being exotic for the ordinary Einstein-Hilbert gravity. Specifically we provide the vacuum unimodular mimetic gravity description of the de Sitter cosmology and of the perfect fluid with constant equation of state cosmology. As we demonstrate, these cosmologies can be realized by vacuum mimetic unimodular gravity, without the existence of any matter fluid source. Moreover, we investigate how cosmologically viable cosmologies, which are compatible with the recent observational data, can be realized by the vacuum unimodular mimetic gravity. Since in some cases, the graceful exit from inflation problem might exist, we provide a qualitative description of the mechanism that can potentially generate the graceful exit from inflation in these theories, by searching for the unstable de Sitter solutions in the context of unimodular mimetic theories of gravity.
\end{abstract}


\maketitle

\section{Introduction}

The unimodular gravity approach \cite{Alvarez:2006uu,Ellis:2010uc,Kluson:2014esa,Barcelo:2014mua,
Burger:2015kie,Alvarez:2015sba,Jain:2012gc,Jain:2011jc,Cho:2014taa,Basak:2015swx,
Gao:2014nia,Nojiri:2015sfd,Nojiri:2016ygo,Nojiri:2016mlb}, offers a consistent theoretical framework, in the context of which, one of the most intriguing problems in theoretical physics and cosmology, finds an elegant solution, namely the cosmological constant problem \cite{Peebles:2002gy}. Particularly, the quantum field theory estimations of the vacuum energy originating from the vacuum expectation value of a scalar field cosmological constant, is $60$-$120$ orders higher in magnitude, when compared to the observed value of the cosmological constant. The unimodular gravity framework, however, makes it possible to generate a cosmological constant by using the corresponding mechanism, in a natural and intrinsic to the theory way, in which case the value of the cosmological constant can suitably be adjusted to the observed value. Specifically, the cosmological constant originates from the trace-free part of the resulting Einstein field equations, with the trace-free part being obtained by employing the constraint that the metric determinant $\sqrt{-g}$ is a fixed function of spacetime, or a constant number. Apart from the cosmological constant, the unimodular gravity theoretical framework can harbor another conceptually intriguing feature of the up to present date observed Universe, the late-time acceleration (see for example Refs.~\cite{Jain:2012gc,Jain:2011jc,Cho:2014taa}), firstly observed in the late 90's \cite{Nojiri:2016mlb}. Moreover and interestingly enough, the cosmological perturbations in the standard Einstein-Hilbert gravity and in unimodular gravity are quantitatively the same, when the linear perturbation theory is used, as was explicitly demonstrated in Refs.~\cite{Basak:2015swx,Gao:2014nia}, with some differences occurring however in the Sachs-Wolfe relation between gravitational potential and microwave temperature anisotropies, see \cite{Gao:2014nia} for more details on this issue. Recently, as an extension of the unimodular Einstein-Hilbert general relativity, the unimodular $F(R)$ gravity was proposed, see \cite{Nojiri:2015sfd,Nojiri:2016ygo}.


Since the unimodular gravity framework can potentially explain the cosmological constant problem and the late-time acceleration of the Universe, it would be interesting to combine the unimodular gravity theory with the mimetic gravity approach, firstly developed in \cite{Chamseddine:2013kea,Chamseddine:2014vna}, later generalized as mimetic $F(R)$ gravity proposed in \cite{Nojiri:2014zqa}. The cosmology of extended mimetic gravity (first of all, $F(R)$ mimetic gravity) was investigated in Refs.~\cite{Chamseddine:2014vna,Astashenok:2015haa,Momeni:2015gka,Raza:2015kha,Rabochaya:2015haa, Myrzakulov:2015qaa,Astashenok:2015qzw,Myrzakulov:2015nqa, Mirzagholi:2014ifa,Oikonomou:2015lgy,Cognola:2016gjy,Hammer:2015pcx,Nojiri:2014zqa,Odintsov:2015cwa,Odintsov:2015wwp,Odintsov:2015ocy}. The reason to attempt this kind of theoretical unification of mimetic and unimodular gravity, is that mimetic gravity can consistently explain dark matter in a geometrical manner, without the need of using cold dark matter to address this issue. The dark matter issue is one of the great questions posed to theoretical physicists, even up to date, and it is believed to govern the evolution of our Universe in a percentage of $26.8\%$, since the corresponding density of the Universe is $\Omega_\mathrm{DM}\sim 26.8\%$. In the literature there exist many possible candidates for dark matter, which describe dark matter as a particle \cite{Oikonomou:2006mh,Shafi:2015lfa}, but since no clear verification for this claim exist, all possible theories should be taken seriously and studied thoroughly. Actually, in the context of mimetic $F(R)$ gravity, late and early-time acceleration can be explained, as was demonstrated in Refs.~\cite{Nojiri:2014zqa,Odintsov:2015cwa,Odintsov:2015wwp,Odintsov:2015ocy}, and also it is possible to achieve concordance with the recent Planck \cite{Ade:2015lrj,Planck:2013jfk} and BICEP2/Keck-Array data \cite{Array:2015xqh}, as was explicitly demonstrated in \cite{Odintsov:2015cwa,Odintsov:2015wwp,Odintsov:2015ocy}.

Conceptually, in order to combine the two theories and to solve in a unified geometrical way, the cosmological constant problem and the dark matter issue, we shall use the Lagrange multiplier approach \cite{Lim:2010yk,Capozziello:2013xn,Capozziello:2010uv}, in which case, the unimodular and mimetic constraints shall be realized by introducing in the action some appropriately chosen Lagrange multipliers. Then, by varying the action with respect to these Lagrange multipliers, the unimodular and mimetic constraints naturally arise in the theory as parts of the equations of motion. We need to note that the unimodular constraint is just a constraint on the allowed metric in the theory, while the mimetic approach makes use of the internal conformal degrees of freedom of the metric, with these being quantified in terms of a scalar field $\phi$, which we call mimetic scalar field hereafter. This scalar field is an auxiliary degree of freedom, with only its first derivatives appearing in the action. Consequently, in order to keep the two theoretical concepts clear in our unified mimetic unimodular approach, we will make use of the Lagrange multipliers method, since this will keep the two concepts clear and it is more easy to extract the physical information from the gravitation action. The mimetic theory we shall use will be inspired by the mimetic theory of Ref.~\cite{Chamseddine:2013kea} in which case, a potential term  of the auxiliary scalar field $V(\phi)$ and a Lagrange multiplier was used too. 
  
The purpose of this paper is to introduce the theoretical framework of unimodular-mimetic (shortened to U-M hereafter) gravity and study some simple cosmological implications of this theory. Particularly, we will generalize mimetic gravity with Lagrange multipliers action, by introducing a unimodular Lagrange multiplier, in terms of which, the unimodular constraint on the metric, shall be realized. After discussing some fundamental issues about the U-M gravity approach, we describe in detail the reconstruction method underlying the formalism. By using this reconstruction method we will realize various cosmological scenarios, some of which however, being exotic for the standard Einstein-Hilbert gravity. Hence, as we demonstrate, the U-M gravity proves to be a useful tool for the realization of various cosmologies. Also, in order to further support the utility of our results, 
We will use the perfect fluid approach developed in Ref.~\cite{Bamba:2014wda}, aiming in realizing cosmologies compatible with the recent observational data of Planck \cite{Ade:2015lrj,Planck:2013jfk} and BICEP2/Keck-Array \cite{Array:2015xqh}. We will be interested in realizing inflationary models \cite{Mukhanov:1990me,Gorbunov:2011zzc,Linde:2014nna,Bamba:2015uma,Lyth:1998xn,Miao:2015oba}, for which we compute the spectral index of primordial curvature perturbations \cite{Mukhanov:1990me,Brandenberger:1983tg,Brandenberger:1983vj,Baumann:2009ds} and also the scalar-to-tensor index. As we demonstrate, concordance with both the Planck \cite{Ade:2015lrj,Planck:2013jfk} and BICEP2/Keck-Array \cite{Array:2015xqh} data can be achieved. Finally, we investigate how graceful exit can be achieved in the context of U-M gravity, and as we show, graceful exit can occur if the unstable de Sitter solution exists in the U-M theory \cite{Bamba:2014jia}. Then graceful exit can be triggered by unstable curvature perturbations which grow in time \cite{Bamba:2014jia}. In fact, this is an alternative mechanism to the standard slow-roll approach in scalar-tensor theories, although some qualitative difference exists between these two approaches, since in both cases the de Sitter final attractor ceases to be the final attractor solution. In the unstable curvature perturbations case, this is owing to the fact that the unstable de Sitter solution occurs, while in the slow-roll case, this is owing to the breakdown of the perturbative slow-roll expansion (for a thorough and informative account on the slow-roll expansion, see \cite{Liddle:1994dx,Liddle:1992wi,Copeland:1993jj}).

We need to stress that the findings of this paper do not provide the ultimate solution to the cosmological constant and dark matter issue. Our purpose is to provide a new suggestive proposal for a theoretical framework that may assist in the discussion and solution search of two significant problems, and of course our framework is not the ultimate theory, but just another solution that provides results compatible with observations. Also our framework makes possible the realization of two cosmologies, in the absence of any matter fluids or cosmological constant. Particularly, in the case of the standard Einstein-Hilbert gravity, the de Sitter cosmology could be realized only by the presence of a cosmological constant and the perfect fluid cosmology, in which case the scale factor as a function of the cosmic time behaves as $a(t)\sim t^{2/(3(1+w))}$, could be realized if a perfect fluid with energy density $\rho\sim a^{-3(1+w)}$ was present. In the mimetic unimodular case, these two cosmologies can be realized by the vacuum theory, without any matter fluids being present. This is one novelty of the present formalism, which however also occurred in the mimetic gravity theory, in which case dark matter occurred as an outcome of the hidden conformal degrees of freedom of the metric, and hence it has a purely geometric origin and no source generated such an evolution. Hence our formalism is aligned more or less with the mimetic gravity line of research,  which generates cosmologies that were realized in the standard Einstein-Hilbert gravity only if matter fluid sources were present.   

Naturally by combining the mimetic and unimodular formalisms, one ends up in two constraint equations, one of which is non-covariant, namely the one corresponding to the unimodular constraint. In order to remedy this issue, we also provide an alternative covariant version of our theoretical framework and we discuss how the equations of motion of the resulting theory are affected.

This paper is organized as follows: In section II we present the unimodular mimetic gravity with potential formalism and explain in detail how the underlying reconstruction technique works. In section III we demonstrate how some quite well known cosmologies can be realized from the vacuum theory of unimodular mimetic gravity, and we compare our findings to the mimetic gravity results in order to see the differences and new insights that the unimodular mimetic gravity approach brings along. In section IV we use the perfect fluid approach in order to calculate the observational indices for some inflationary models which are in concordance with the observational data, and we investigate how these cosmologies can be realized in the context of unimodular mimetic gravity theory. In section V we address the graceful exit issue by investigating whether the unstable de Sitter solutions exist in the U-M gravity theory. Finally, in section VI we provide a covariant version of the mimetic unimodular theory. The conclusions follow at the end of the paper.

\section{The Unimodular-Mimetic Gravity Formalism}

In this section we present the formalism of U-M gravity, which will enable us to realize various cosmological scenarios. This formalism constitutes a reconstruction method, and we now present the essential features of this method. The standard Einstein-Hilbert gravity approach \cite{Alvarez:2006uu,Ellis:2010uc,Kluson:2014esa,Barcelo:2014mua,
Burger:2015kie,Alvarez:2015sba,Jain:2012gc,Jain:2011jc,Cho:2014taa,Basak:2015swx,Gao:2014nia}, is based on the assumption that the determinant of the metric tensor is fixed, so that the metric satisfies $g_{\mu \nu}\delta g^{\mu \nu}=0$, which implies that the components of the metric tensor can be adjusted in such a way, so that the determinant of the metric $\sqrt{-g}$ is a fixed function of spacetime, in the most general case. Hence, hereafter we assume that the metric satisfies,  
\be
\label{Uni1}
\sqrt{-g}=1\, ,
\ee
to which we shall refer to as the ``unimodular constraint''. In order to realize the unimodular constraint, we shall use the Lagrange multiplier method, so the constraint appears as part of the corresponding equations of motion. Having this in mind, a direct generalization of the mimetic gravity with potential $V(\phi )$ and Lagrange multiplier $\eta (\phi)$ of Ref.~\cite{Chamseddine:2014vna}, that takes into account the unimodular constraint of Eq.~(\ref{Uni1}), is the following,
\be
\label{UM1}
S = \int d^4 x \left\{ \sqrt{-g} \left( \frac{R}{2\kappa^2}+f(R) - V(\phi) 
 - \eta \left( \partial_\mu \phi \partial^\mu \phi  + 1 \right)
 - \lambda \right) + \lambda \right\} 
+ S_\mathrm{matter} \, ,
\ee
where $\phi$ is the real mimetic scalar field, $R$ denotes as usual the Ricci scalar, and also $S_\mathrm{matter}$ represents the action for the matter fields present. Note that the action (\ref{UM1}) describes the $F(R)=R+f(R)$ gravitational action, but in this paper for simplicity we take $f(R)=0$, and the $F(R)$ case will be studied elsewhere. In addition, the functions $\eta$ and $\lambda$ are the Lagrange multiplier fields, with $\eta$ being the one directly related to the mimetic gravity, while $\lambda$ is introduced in order the unimodular constraint is realized. Indeed, upon variation of the action of Eq.~ (\ref{UM1}), with respect to the function $\eta$, we obtain the following constraint,
\be
\label{UM2}
\partial_\mu \phi \partial^\mu \phi =  - 1 \, ,
\ee
which is the mimetic constraint, also found in \cite{Chamseddine:2013kea,Chamseddine:2014vna}. In addition, by varying the action (\ref{UM1}), with respect to the function $\lambda$ this time, we easily obtain the unimodular constraint of Eq.~(\ref{Uni1}). As we demonstrate shortly, both the functions $\eta $ and $\lambda$ are functions of the cosmic time $t$, which will be identified to the auxiliary scalar field $\phi$, which holds true owing to the mimetic constraint and also by assuming that the auxiliary field is a function of the cosmic time $t$. The resulting Einstein field equations can be obtained by varying the action of Eq.~(\ref{UM1}) with respect to the metric, and we obtain the following sets of equations,
\be
\label{UM2B}
0=\frac{1}{2}g_{\mu\nu} \left( \frac{R}{2\kappa^2} - V(\phi) 
 - \eta \left( \partial_\mu \phi \partial^\mu \phi  + 1 \right)
 - \lambda \right)  - \frac{1}{2\kappa^2} R_{\mu\nu} 
+ \eta \partial_\mu \phi \partial_\nu \phi + \frac{1}{2} T_{\mu\nu} \, ,
\ee
with $T_{\mu\nu}$ denoting the energy-momentum tensor of the perfect matter fluids present. Also, by varying the action (\ref{UM1}) with respect to the auxiliary scalar field $\phi$, yields the following equation, 
\be
\label{UM3}
0 = 2 \nabla^\mu \left( \lambda \partial_\mu \phi \right) - V' (\phi)\, .
\ee
We assume that the background metric is a flat Friedman-Robertson-Walker (FRW) metric with the line element being of the form,
\be
\label{FRW}
ds^2 = - dt^2 + a(t)^2 \sum_{i=1}^3 \left( dx^i \right)^2 \, .
\ee
Owing to the fact that this metric does not satisfy the unimodular constraint of Eq.~(\ref{Uni1}), we redefine the cosmic time variable $t$ of the FRW metric, in such a way so that the unimodular constraint (\ref{Uni1}) is satisfied, as follows, $ d\tau = a(t)^3  dt$.
Correspondingly, the FRW metric of Eq.~(\ref{FRW}) can be rewritten in the following way,
\be
\label{UniFRW}
ds^2 = - a\left(t\left(\tau\right)\right)^{-6} d\tau^2 + a\left(t\left(\tau\right)\right)^{2} 
\sum_{i=1}^3 \left( dx^i \right)^2 \, .
\ee
In the following we shall refer to the metric of Eq.~(\ref{UniFRW}) as the unimodular FRW metric. For the unimodular metric Eq.~(\ref{UniFRW}), the non-vanishing components of the Levi-Civita connection and of the Ricci tensor are given by,
\begin{align}
\label{Uni13}
& \Gamma^t_{tt} = - 3 K\, , \quad \Gamma^t_{ij} = a^8 K \delta_{ij}\, , \quad 
\Gamma^i_{jt} = \Gamma^i_{tj} = K \delta_j^{\ i} \, , \nn
& R_{tt} = - 3 \frac{d K}{d\tau} - 12 K^2\, , \quad 
R_{ij} = a^8 \left( \frac{d K}{d\tau} + 6 K^2 \right) \delta_{ij}\, .
\end{align}
In Eq.~(\ref{Uni13}), the function of $\tau$ is a direct generalization of the Hubble rate in terms of the $\tau$ variable, that is, $K\equiv \frac{1}{a} \frac{da}{d\tau}$. Moreover, the Ricci scalar $R$ in terms of the $\tau$ variable is equal to,
\begin{equation}
\label{scalarunifrw}
R = a^6 \left( 6 \frac{d K}{d\tau} + 30 K^2 \right) \, .
\end{equation}
Since we assumed that the auxiliary scalar $\phi$ depends on the time coordinate $t$ (or equivalently $\tau$), the mimetic constraint of Eq.~(\ref{UM2}) can be cast in the following way,
\be
\label{UM4}
a^{-6}\left( \frac{d\phi}{d\tau} \right)^2 = 1\, ,
\ee
which can be rewritten in terms of the cosmological time $t$ by using $ d\tau = a(t)^3  dt$, as follows,
\be
\label{UM5}
\left( \frac{d\phi}{dt} \right)^2  = 1\, .
\ee
Hence we may identify the auxiliary field $\phi$ with the cosmological time $t$, that is $\phi=t$. The $(\tau,\tau)$ and $(i,j)$ components of the equations (\ref{UM2B}) yield the following equations,
\begin{align}
\label{UM6}
0 =& - \frac{3a^6}{2\kappa^2} K^2 + \frac{V(\phi)}{2} + \frac{\lambda}{2} + \eta + \frac{\rho}{2}\, , \\
\label{UM7}
0 =&  \frac{a^6}{2\kappa^2} \left( 2 \frac{dK}{d\tau} + 9 K^2\right)  - \frac{V(\phi)}{2} - \frac{\lambda}{2} + \frac{p}{2}\, ,
\end{align}
where $\rho$ and $p$ denote the energy density and the pressure of the matter fluids present. Note that in order to obtain Eqs.~(\ref{UM6}) and (\ref{UM7}), we made use of the constraint of Eq.~(\ref{UM4}). We can rewrite Eqs.~(\ref{UM6}) and (\ref{UM7}) by using the cosmological time $t$ variable, in the following way,
\begin{align}
\label{UM8}
0 = & - \frac{3 H^2}{2\kappa^2} + \frac{V(\phi)}{2} + \frac{\lambda}{2} + \eta + \frac{\rho}{2}\, , \\
\label{UM9}
0 = & \frac{1}{2\kappa^2} \left( 3 H^2 + 2 \frac{dH}{dt} \right) - \frac{V(\phi)}{2} - \frac{\lambda}{2} + \frac{p}{2}\, .
\end{align}
Moreover, Eq.~(\ref{UM3}) can be written as follows,
\be
\label{UM10}
0 = - 6H \lambda - 2 \frac{d\lambda}{dt} - V'(\phi)\, .
\ee
Then by making use of Eqs.~(\ref{UM9}) and (\ref{UM10}), we may delete $\lambda$ from the equations of motion, and we obtain,
\be
\label{UM11}
0 = 6 H V(\phi) - 3 V'(\phi) - 6 H p - 2 \frac{dp}{dt} 
+ \frac{1}{\kappa^2} \left( - 18 H^3 - 6 H \frac{dH}{dt} + 4 \frac{d^2 H}{dt^2} \right)\, .
\ee
Since the auxiliary scalar field is identified with the cosmic time $\phi = t$, we can easily integrate Eq.~(\ref{UM12}) and obtain the potential $V(\phi)$,
\be
\label{UM12} 
V(\phi) = \frac{a \left( t = \phi \right)^2}{3} \int^\phi dt \, a(t)^{-2} 
\left\{ - 6 H(t) p(t) - 2 \frac{dp(t)}{dt} + \frac{1}{\kappa^2} \left( - 18 H(t)^3 
 - 6 H(t) \frac{d H(t)}{dt} + 4 \frac{d^2 H(t)}{dt^2} \right) \right\} \, .
\ee
Hence, for a specific cosmology with scale factor $a(t)$, and if the time dependence of the energy density and pressure in terms of the scale factor are given, by using Eq.~(\ref{UM12}) we can obtain the exact form of the mimetic potential $V(\phi=t)$, which can realize the cosmological evolution with scale factor $a(t)$. Having that at hand, by using Eqs.~(\ref{UM9}) and (\ref{UM8}), we may solve with respect to the functions $\lambda (t)$ and $\eta (t)$, and therefore determine completely the unimodular mimetic gravity with potential that realizes the cosmological evolution with scale factor $a(t)$. Basically, the resulting Eqs.~(\ref{UM12}), (\ref{UM9}) and (\ref{UM8}), constitute the unimodular mimetic reconstruction method, and in the next sections we will make extensive use of these, in order to realize various well known cosmological scenarios. Before proceeding to the examples, in the next section we shall demonstrate that the unimodular mimetic gravity can be described by a perfect fluid.

\subsection{Perfect Fluid Description of the Unimodular-Mimetic Gravity}

As we now demonstrate, it is possible to describe the effect of unimodular mimetic gravity with potential in terms of a perfect fluid. This was also possible in mimetic gravity with scalar potential, as was shown in Ref.~\cite{Chamseddine:2014vna}. We start off with taking the trace of the equation of motion appearing in Eq.~(\ref{UM2B}), and the result is,
\begin{equation}\label{etavar}
\eta (t)=\frac{1}{\kappa^2}\left( G-4\kappa^2 \tilde{V}(t)-\kappa^2\right)\, ,
\end{equation}
where $G$ and $T$ stand for the trace of the Einstein tensor and of the energy momentum tensor respectively. In addition, we introduced the potential term $\tilde{V}$, which is equal to,
\begin{equation}\label{eksed}
\tilde{V}(t)=-\lambda (t)-V(t)\, .
\end{equation}
The perfect fluid describing the unimodular mimetic gravity is described by the energy density $\tilde{\rho}$ and the effective pressure $\tilde{p}$, which are given below,
\begin{equation}\label{energyandpressure}
\tilde{\rho}=G-\kappa^2 T-4\kappa^2\tilde{V}\, ,\quad \tilde{p}=-\tilde{V}\kappa^2\, .
\end{equation}
As we now demonstrate, the effective energy density and effective pressure satisfy a conservation law. By acting on Eq.~(\ref{UM2B}) with the covariant derivative $\nabla^{\nu}$ and also by taking into account that $\nabla^{\rho}\left( g^{\mu \nu}\partial_{\mu}\phi \partial_{\nu}\phi \right)=2g^{\mu \nu}(\nabla_{\mu}\partial^{\rho}\phi )\partial_{\nu}\phi=0$ (see also \cite{Chamseddine:2014vna}), we obtain the following equation,
\begin{equation}\label{followup}
\nabla^{\mu}\left( (G-\kappa^2 T-4\kappa^2\tilde{V})\partial_{\nu} \phi\right)=-\kappa^2\dot{\tilde{V}}\, .
\end{equation}
Since the Christoffel symbols $\Gamma^{2}_{2\,1}$, $\Gamma^{3}_{3\,1}$ and $\Gamma^{4}_{4\,1}$ for the FRW metric of Eq.~(\ref{FRW}) are equal to,
\begin{equation}\label{chris}
\Gamma^{2}_{2\,1}=\Gamma^{3}_{3\,1}=\Gamma^{4}_{4\,1}=\frac{\dot{a}}{a}\, ,
\end{equation}
and by using the definitions of Eq.~(\ref{energyandpressure}), then Eq.~(\ref{followup}) takes the following form,
\begin{equation}\label{conteqn}
\dot{\tilde{\rho } }+3\,H\,\left(\tilde{\rho}+\tilde{p} \right)=0\, ,
\end{equation}
which describes a perfect fluid with the energy density and pressure being functions of the mimetic potential $V(t)$ and of the unimodular Lagrange multiplier $\lambda (t)$, as these appear in Eq.~(\ref{energyandpressure}). In the following sections we demonstrate how various well known cosmological scenarios can be realized by U-M gravity.

\section{Cosmological Evolution with Vacuum Unimodular-Mimetic Gravity}

Having the reconstruction method for U-M gravity at hand, in this section we demonstrate how certain cosmological scenarios can be realized by using the U-M formalism. Note that in some cases, these scenarios were in some sense exotic for the standard Einstein Hilbert gravity, since these could not be realized by the known matter fluids. However, as we demonstrate in this section, even the vacuum U-M theory can consistently and elegantly describe these scenarios. Throughout this section we shall assume that no matter fluid is present, therefore we consider the vacuum U-M theory. We start our analysis by studying the de Sitter cosmology.

Before we start we need to discuss the main features of the theoretical framework we propose. In the standard Einstein-Hilbert gravity, the de Sitter solution could be realized only by the presence of a cosmological constant and also a perfect fluid cosmology, in which case the scale factor as a function of time behaves as $a(t)\sim t^{2/(3(1+w))}$, could only be realized if a perfect fluid with constant equation of state parameter $w$, was present, in which case the energy density as a function of the scale factor is $\rho\sim a^{-3(1+w)}$. The novel feature of the formalism we propose is that these two cosmologies can be realized by the vacuum theory without the presence of perfect fluids. This feature is also a feature of the mimetic theory, since dark matter appears in the resulting FRW equations of motion, without a matter fluid with $w=0$ being present, so dark matter has a purely geometric origin. Hence the framework we propose combines two geometric theories and enables us to realize various cosmologies without the need of any matter fluids. The presence of the potential and of the Lagrange multipliers is what makes the theory able to realize these cosmologies. 

With regards to the potential, in the mimetic gravity case the choice of the potential plays a crucial role in the realization of a specific cosmology, see for example Ref. \cite{Chamseddine:2014vna}. As was demonstrated in Ref. \cite{Chamseddine:2014vna}, the potential $V(t)\sim t^{-2}$ produces a quintessence cosmology, see \cite{Chamseddine:2014vna} for details. In our case, the mimetic unimodular framework provides more freedom owing to the presence of the unimodular and mimetic Lagrange multipliers. Hence, even if the potential is fixed, by appropriately choosing the unimodular and mimetic Lagrange multipliers, a specific cosmology can be realized. Of course the potential and also the two Lagrange multipliers are constrained by the equations of motion, but still there is more freedom in comparison to the mimetic gravity or unimodular gravity. This was our initial motivation for working this theory in the first place, however by no means should our theoretical proposal be considered as the ultimate theory of everything for cosmology. It is just another successful modified gravity description, which provides an alternative approach to cosmological problems. Having discussed these issues, let us now demonstrate how easily two well known cosmologies can be realized in the context of unimodular mimetic gravity.

\subsection{Realization of de Sitter Cosmology from vacuum U-M Gravity}

Consider that the Universe is described by the de Sitter cosmological evolution, in which case the scale factor and the Hubble rate in terms of the cosmic time $t$, are equal to,
\begin{equation}\label{desitterscale}
a(t)=\e^{H_0\, t}\, ,\quad H(t)=H_0
\end{equation}
with $H_0$ being some positive real parameter. Then, by taking into account that we assumed the vacuum U-M theory, the energy density and pressure are equal to zero, that is $\rho(t)=p(t)=0$, and therefore, by using Eq.~(\ref{UM12}), the mimetic potential $V(t)$ reads,
\begin{equation}\label{mimeticpotdesitter}
V(t=\phi)= -\frac{6 \e^{2 H_0 t} H_0^3 t}{\kappa ^2}\, ,
\end{equation}
and hence, by substituting in Eq.~(\ref{UM9}), the unimodular function $\lambda (t)$ can be easily found, and the resulting expression for it is,
\begin{equation}\label{unilamdadesitter}
\lambda (t)=\frac{3 H_0^2 \left(1+2 \e^{2 H_0 t} H_0 t\right)}{\kappa ^2}\, .
\end{equation}
Correspondingly, by substituting the resulting expressions for the Lagrange multiplier function $\lambda (t)$ and for the mimetic potential $V(t)$ from Eqs.~(\ref{mimeticpotdesitter}) and (\ref{unilamdadesitter}) into Eq.~(\ref{UM8}), we can easily obtain the mimetic Lagrange multiplier function $\eta (t)$, which reads,
\begin{equation}\label{mimeticlagrangedesitter}
\eta (t)=\frac{3 H_0^2}{2 \kappa ^2}\, ,
\end{equation}
so the mimetic Lagrange multiplier is a positive constant number for the de Sitter cosmology. A direct comparison of the resulting picture in the context of U-M gravity, with the standard Einstein-Hilbert gravity and also with ordinary mimetic gravity \cite{Chamseddine:2014vna} shows that the U-M gravity result is different, as was probably expected.

In the present description the presence of a cosmological constant is not required by the theory, since the Lagrange multipliers and the potential appropriately realize the de Sitter cosmology.

\subsection{Perfect Fluid Cosmology with Constant Equation of State from U-M Gravity}

In the context of the ordinary Einstein-Hilbert gravity, if a perfect fluid with equation of state $p=w\rho$ is present, then the resulting scale factor and the corresponding Hubble rate are equal to,
\begin{equation}\label{desitterscale1}
a(t)=t^{\frac{2}{3 (1+w)}}\, ,\quad H(t)=\frac{2}{3 t (1+w)}\, ,
\end{equation}
with $w$ being the constant equation of state parameter. We now demonstrate how to realize the cosmological evolution described by (\ref{desitterscale1}), by using the vacuum U-M gravity theory. By using Eq.~(\ref{UM12}), and substituting the scale factor $a(t)$ and the Hubble rate $H(t)$ from Eq.~(\ref{desitterscale1}), the mimetic potential in this case reads,
\begin{equation}\label{mimeticpotdesitter1}
V(t=\phi)= -\frac{4 t^{-2+\frac{4}{3 (1+w)}} \left(1+5 w+2 w^2\right)}{9 (1+w)^3 \kappa ^2}\, ,
\end{equation}
and therefore, by substituting in Eq.~(\ref{UM9}), the unimodular Lagrange multiplier function $\lambda (t)$ in this case is equal to,
\begin{equation}\label{unilamdadesitter1}
\lambda (t)=\frac{4 \left(-3 w (1+w)+t^{\frac{4}{3 (1+w)}} \left(1+5 w+2 w^2\right)\right)}{9 t^2 (1+w)^3 \kappa ^2}\, .
\end{equation}
Finally, by using Eqs.~(\ref{mimeticpotdesitter1}) and (\ref{unilamdadesitter1}), we can easily compute the mimetic Lagrange multiplier function $\eta (t)$ of Eq.~(\ref{UM8}), which reads,
\begin{equation}\label{mimeticlagrangedesitter1}
\eta (t)=\frac{2 (2+w)}{3 t^2 (1+w)^2 \kappa ^2}\, .
\end{equation}
The cosmology described by the scale factor and the Hubble rate of Eq.~(\ref{desitterscale1}), was also studied in the context of ordinary mimetic gravity with potential in Ref.~\cite{Chamseddine:2014vna}, and the mimetic potential which generated this cosmology was found to be equal to,
\begin{equation}\label{mimeperfectfluidcase}
V(t)=\frac{\mathcal{C}}{t^2}\, ,
\end{equation}
with $\mathcal{C}$ some arbitrary constant parameter. Notice that for the case that the equation of state parameter is $w\to \infty$, the mimetic potential of the U-M gravity given in Eq.~(\ref{mimeticpotdesitter1}), becomes approximately equal to, $V(t)\sim t^{-2}$, so in this case there might be overlap with the ordinary mimetic gravity result. In general, however, as in the de Sitter case, the resulting physical picture of the U-M gravity theory is different in comparison to the ordinary mimetic gravity. This is a new feature of our theoretical construction, which we need to report.

Also, note that the presence of a perfect fluid is not necessary in order to produce the perfect fluid cosmology of Eq. (\ref{desitterscale1}). In the standard Einstein-Hilbert gravity, the cosmology (\ref{desitterscale1}), could be realized if a perfect fluid with energy density $\rho\sim a^{-3(1+w)}$ and pressure $p=w\rho$ was present, but in our case this cosmology is realized only by the vacuum theory. This behavior is also a feature of the mimetic gravity with Lagrange multiplier approach, hence our framework provides an alternative to the unimodular and mimetic theories.

\section{Unimodular-Mimetic Gravity Slow-roll Inflation: The Perfect Fluid Approach}

As we demonstrated in the previous sections, the U-M gravity theoretical framework makes possible the realization of various cosmological scenarios, some of which were rather exotic for the ordinary Einstein-Hilbert gravity. In this section we demonstrate that it is possible to realize cosmologies which are compatible with the observational data. Specifically, we demonstrate that compatibility with the latest Planck data \cite{Ade:2015lrj,Planck:2013jfk} and with the recent BICEP2/Keck-Array data \cite{Array:2015xqh} can be achieved. Note that according to the recent Planck data \cite{Ade:2015lrj,Planck:2013jfk} the spectral index $n_s$ and the scalar-to-tensor ratio, satisfy the following constraints,
\begin{equation}\label{constraintedvalues}
n_s=0.9644\pm 0.0049\, , \quad r<0.10\, ,
\end{equation}
while according to the recent BICEP2/Keck-Array data \cite{Array:2015xqh}, the scalar to tensor ratio is further constrained to satisfy 
\begin{equation}\label{bicep2keckarray}
r<0.07\, .
\end{equation}
In order to calculate the observational indices corresponding to the power spectrum of primordial curvature perturbations and to the scalar-to-tensor ratio, we will make use of a very useful technique that treats the modified gravity as a perfect fluid, which was firstly developed in Ref.~\cite{Bamba:2014wda}. This technique holds true when slow-roll evolution takes place, and as was shown in Ref.~\cite{Bamba:2014wda} (see also \cite{Mukhanov:2014uwa}), if the slow-roll condition is satisfied, both the slow-roll indices and the observational indices can be written in terms of the Hubble rate in a convenient and model-independent way. We assume that the Hubble rate is a function of the $e$-foldings number $N$, which is related to the scale factor $a(t)$ as follows, $\e^N=a/a_0$, with $a_0$ the value of the scale factor at some initial time instance. Then, following Ref.~\cite{Bamba:2014wda}, given the cosmological evolution $H(N)$, the slow-roll inflationary indices $\epsilon$, $\eta$, are expressed in terms of the Hubble rate $H(N)$ as follows,
\begin{align}
\label{S7}
\epsilon
=& - \frac{ H(N)}{4  H'(N)} \left( \frac{6\frac{ 
H'(N)}{ H(N)}
+ \frac{ H''(N)}{ H(N)} + \left( \frac{ H'(N)}{ 
H(N)} \right)^2}
{3 + \frac{ H'(N)}{ H(N)}} \right)^2 \, , \nn
\eta = & -\frac{1}{2} \left( 3 + \frac{ H'(N)}{ H(N)} \right)^{-1} 
\left(
9 \frac{ H'(N)}{ H(N)} + 3 \frac{ H''(N)}{ H(N)}
+ \frac{1}{2} \left( \frac{ H'(N)}{ H(N)} \right)^2 -\frac{1}{2} 
\left( \frac{ H''(N)}{ H'(N)} \right)^2
+ 3 \frac{ H''(N)}{ H'(N)} + \frac{ H'''(N)}{ H'(N)} 
\right) \, . 
\end{align}
As was demonstrated in Ref.~\cite{Bamba:2014wda}, in the context of the perfect fluid approach, the spectral index of primordial curvature perturbations $n_s$ can be written in terms of the slow-roll parameters of Eq.~(\ref{S7}), in the following way,
\begin{equation}\label{indexspectrscratio}
n_s\simeq 1-6 \epsilon +2\eta\, ,
\end{equation}
while the scalar to tensor ratio $r$ is equal to,
\begin{equation}\label{stotensration}
\quad r=16\epsilon \, ,
\end{equation}
with both the relations (\ref{indexspectrscratio}) and (\ref{stotensration}) holding true when the slow-roll indices of Eq.~(\ref{S7}) satisfy the constraint $\epsilon,\eta \ll 1$, with the latter constraint materializing the slow-roll approximation.

\begin{figure}[h] \centering
\includegraphics[width=15pc]{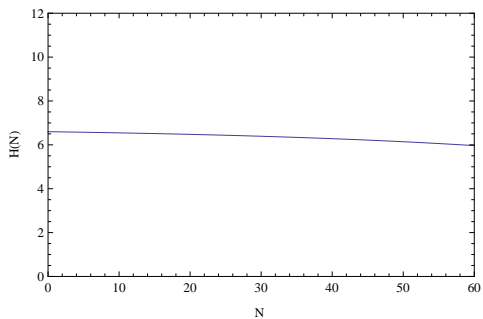}
\caption{The Hubble rate $H(N)$ as a function of the $e$-foldings number $N$, for the cosmological evolution $H(N)=\left(-\alpha\, \e^{\beta N }+\gamma\right)$, for $\alpha=0.2$, $ \gamma=6.8$ and  $\beta=\frac{1}{42}$.}
\label{mattbounce1}
\end{figure}

Consider the following cosmological evolution,
\begin{equation}\label{hub2}
H(N)=\left(-\alpha\, \e^{\beta N }+\gamma\right)\, .
\end{equation}
with $\alpha$, $\beta$, $\gamma $ positive numbers appropriately chosen so that the Hubble rate does not become negative. Particularly, as we show shortly, a convenient set of values for the parameters $\alpha,\beta$ and $\gamma $,is the following,
\begin{equation}\label{parmchoice12}
\alpha=0.2\, , \quad \gamma=6.8\, , \quad \beta=\frac{1}{42}\, ,
\end{equation}
and also we assume that the $e$-foldings number $N$ takes the values $0\leq N\leq 60$, so practically inflation ends in approximately $60$ $e$- foldings. The question how the inflationary evolution might come to a graceful exit is addressed appropriately in the next section. For the values of the parameters as chosen in Eq.~(\ref{parmchoice12}), in Fig.~\ref{mattbounce1}, we plot the Hubble rate $H(N)$ as a function of $N$, and as it can be seen, it is always positive from $N=60$ up to $N=0$ where inflation is supposed to end. 
By substituting the Hubble rate of Eq.~(\ref{hub2}) into the slow-roll parameters of Eq.~(\ref{S7}), the slow-roll parameters take the following form,
\begin{align}\label{hubslowroll2}
\epsilon=&-\frac{\e^{N \beta } \alpha  \beta  \left(-2 \e^{N \beta } \alpha  (3+\beta )+(6+\beta ) \gamma \right)^2}{4 \left(\e^{N \beta } \alpha  (3+\beta )-3 \gamma \right)^2 \left(\e^{N \beta } \alpha -\gamma \right)} \, ,  \\
\label{edggs}
\eta = & -\frac{\beta  \left(8 \e^{2N \beta } \alpha ^2 (3+\beta )-2 \e^{N \beta } \alpha  (15+4 \beta ) \gamma +(6+\beta ) \gamma ^2\right)}{4 \left(\e^{N \beta } \alpha  (3+\beta )-3 \gamma \right) \left(\e^{N \beta } \alpha -\gamma \right)} \, .
\end{align}
Accordingly, by substituting the slow-roll parameters $\epsilon$ and $\eta$ from Eqs.~(\ref{hubslowroll2}) and (\ref{edggs}), into Eq.~(\ref{indexspectrscratio}), the spectral index $n_s$ becomes equal to,
\begin{align}
\label{scalarpertandsctotenso}
n_s= & \frac{2 \e^{3N \beta } \alpha ^3 (3+\beta )^2 (1+2 \beta )-2 \e^{2N \beta } \alpha ^2 \left(27+39 \beta +16 \beta ^2+2 \beta ^3\right) \gamma }{2 \left(\e^{2N \beta } \alpha  (3+\beta )-3 \gamma \right)^2 \left(\e^{N \beta } \alpha -\gamma \right)} \nn
& +\frac{\e^{2N \beta } \alpha  \left(54+12 \beta +3 \beta ^2+2 \beta ^3\right) \gamma ^2+3 \left(-6+6 \beta +\beta ^2\right) \gamma ^3}{2 \left(\e^{2N \beta } \alpha  (3+\beta )-3 \gamma \right)^2 \left(\e^{N \beta } \alpha -\gamma \right)} 
\, ,
\end{align}
and correspondingly, the scalar-to-tensor ratio $r$ receives the following form,
\begin{equation}\label{thodorakis}
r=-\frac{4 \e^{N \beta } \alpha  \beta  \left(-2 \e^{N \beta } \alpha  (3+\beta )+(6+\beta ) \gamma \right)^2}{\left(\e^{N \beta } \alpha  (3+\beta )-3 \gamma \right)^2 \left(\e^{N \beta } \alpha -\gamma \right)} \, .
\end{equation}
So by substituting the values of the parameters as chosen in Eq.~(\ref{parmchoice12}), the spectral index of primordial curvature perturbations $n_s$ and the scalar-to-tensor ratio $r$ become equal to,
\begin{equation}\label{indnewparadigm12}
n_s\simeq 0.965984\, , \quad r=0.0537777\, ,
\end{equation}
which are in concordance with both the latest Planck data \cite{Ade:2015lrj,Planck:2013jfk} and with the BICEP2/Keck-Array data \cite{Array:2015xqh}, as it can be seen by looking the constraints in Eqs.~(\ref{constraintedvalues}) and (\ref{bicep2keckarray}). 

Now we demonstrate how the viable cosmology of Eq.~(\ref{hub2}) can be realized in the context of U-M gravity. Before getting started, it is more convenient for the reconstruction technique we use, to express the cosmological evolution of Eq.~(\ref{hub2}) in terms of the cosmic time $t$. By using the relations $H=\dot{a}/a$, $\e^N=a/a_0$, the scale factor corresponding to the Hubble rate of Eq.~(\ref{hub2}) is equal to,
\begin{equation}\label{neweqns1}
a(t)=\frac{\e^{(c_1+t) \beta  \gamma } \gamma }{1+a_0^{-\beta } \e^{(c_1+t) \beta  \gamma } \alpha }\, ,
\end{equation}
where $c_1$ is an arbitrary integration constant. By substituting the scale factor and the corresponding Hubble rate in Eq.~(\ref{UM12}), we obtain the mimetic potential which is, 
\begin{equation}\label{mimeticpotdesitter3a1}
V(t=\phi)= -\frac{2 a_0^{\beta } \e^{2 (c_1+t) \beta  \gamma } \beta ^3 \gamma ^5 \left(9 a_0^{-\beta } t+\frac{4 a_0^{\beta }}{\left(a_0^{\beta }+\e^{(c_1+t) \beta  \gamma } \alpha \right)^2 \beta  \gamma }+\frac{11}{a_0^{\beta } \beta  \gamma +\e^{c_1 \beta  \gamma +t \beta  \gamma } \alpha  \beta  \gamma }-\frac{9 a_0^{-\beta } \ln\left(a_0^{\beta }+\e^{(c_1+t) \beta  \gamma } \alpha \right)}{\beta  \gamma }\right)}{3 \left(\kappa +a_0^{-\beta } \e^{(c_1+t) \beta  \gamma } \alpha  \kappa \right)^2}\, .
\end{equation}
Accordingly, by substituting the mimetic potential of Eq.~(\ref{mimeticpotdesitter3a1}) in Eq.~(\ref{UM9}), we obtain the unimodular Lagrange multiplier function $\lambda (t)$, which is, 
\begin{align}\label{unilamdadesittera1}
\lambda (t)=&\frac{1}{3 \left(a_0^{\beta }+\e^{(c_1+t) \beta  \gamma } \alpha \right)^4 \kappa ^2}a_0^{\beta } \beta ^2 \gamma ^2 \left(-6 \e^{3 (c_1+t) \beta  \gamma } \alpha ^3+3 a_0^{\beta } \e^{2 (c_1+t) \beta  \gamma } \alpha ^2 \left(-1+6 \e^{2 (c_1+t) \beta  \gamma } t \beta  \gamma ^3\right)\right. \nn 
& 3 a_0^{3 \beta } \left(3+2 \e^{2 (c_1+t) \beta  \gamma } \gamma ^2 (5+3 t \beta  \gamma )\right)+2 a_0^{2 \beta } \e^{(c_1+t) \beta  \gamma } \alpha  \left(6+\e^{2 (c_1+t) \beta  \gamma } \gamma ^2 (11+18 t \beta  \gamma )\right) \nn 
& \left.-18 a_0^{\beta } \e^{2 (c_1+t) \beta  \gamma } \left(a_0^{\beta }+\e^{(c_1+t) \beta  \gamma } \alpha \right)^2 \gamma ^2 \ln\left(a_0^{\beta }+\e^{(c_1+t) \beta  \gamma } \alpha \right)\right)
\, .
\end{align}
Finally, by combining Eqs.~(\ref{mimeticpotdesitter3a1}), (\ref{unilamdadesittera1}) and (\ref{UM8}), we obtain the mimetic Lagrange multiplier function $\eta (t)$, which is,
\begin{equation}\label{mimeticlagrangedesittera1}
\eta (t)= \frac{a_0^{\beta } \left(3 a_0^{\beta }+2 \e^{(c_1+t) \beta  \gamma } \alpha \right) \beta ^2 \gamma ^2}{2 \left(a_0^{\beta }+\e^{(c_1+t) \beta  \gamma } \alpha \right)^2 \kappa ^2}\, .
\end{equation}
In principle, less complicated expressions for the resulting U-M theory, can be found if the viable cosmology is appropriately chosen. But our aim in this section was to simply demonstrate how a viable cosmology can be described in terms of U-M gravity, and to demonstrate that complicated cosmological evolutions can be realized in the context of U-M gravity. However, a critical issue that should be addressed by every viable inflationary cosmology, is the graceful exit problem, which we address in the next section.

\section{Graceful Exit from Inflation via Unstable de Sitter Solutions}

Apart from being compatible with the observational data, a viable inflationary cosmology should appropriately address the graceful exit issue. Indeed the graceful exit problem, can be a serious issue for many cosmological scenarios, and a consistent solution to this is required, in order for the inflationary cosmology to be considered as a successful description of the Universe at early times. 

\begin{figure}[h] \centering
\includegraphics[width=15pc]{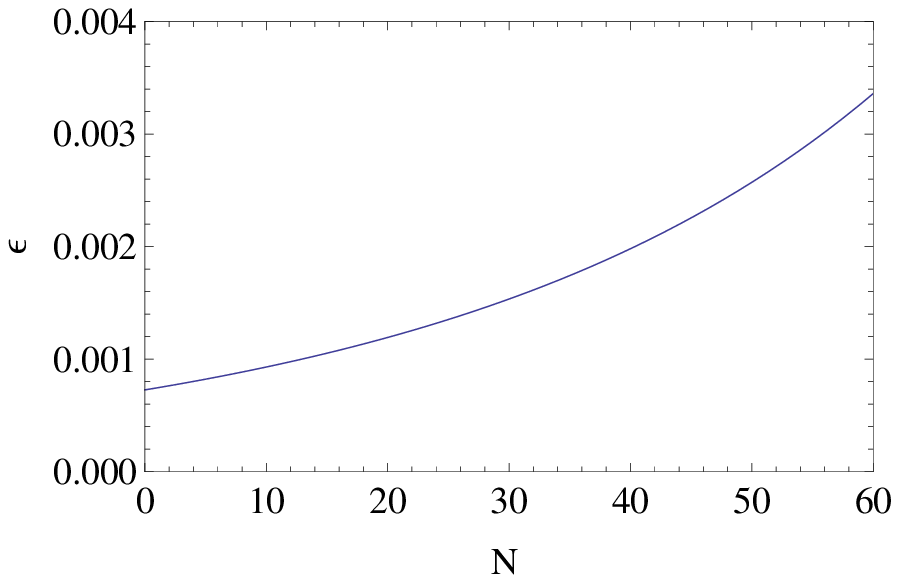}
\includegraphics[width=15pc]{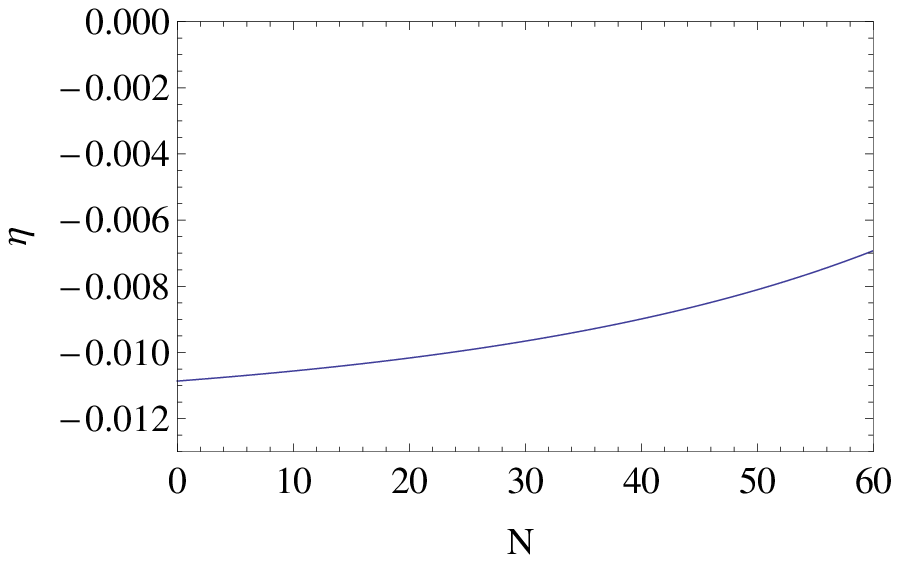}
\caption{The slow-roll parameters $\epsilon$ and $\eta$, for the Hubble rate $H(N)=\left(-\alpha\, \e^{\beta N }+\gamma\right)$, and for $\alpha=0.2$, $ \gamma=6.8$, $\beta=\frac{1}{42}$ and $0\leq N \leq 60$.}
\label{plotslowroll}
\end{figure}

In the standard descriptions of most inflationary cosmologies, which use the slow-roll expansion technique, the graceful exit comes when the slow-roll indices become of order one, and this is sufficient. From another point of view, this description could be insufficient if the slow-roll expansion breaks at a higher order in the slow-roll parameters \cite{Liddle:1994dx,Liddle:1992wi,Copeland:1993jj}. The slow-roll expansion is a powerful tool for determining if the final attractor theorem holds true \cite{Lyth:2009zz}, and the slow-roll parameters $\epsilon$ and $\eta$ are nothing but the first order terms in this expansion. However, there are higher order terms in this expansion which might become of order one, much earlier than the first order slow-roll parameters. In this case, graceful exit occurs at the instance that the slow-roll expansion breaks, even at higher order. Since the manipulation of the full slow-roll expansion for complicated cosmologies, like the one we presented in this section, is rather cumbersome, in this section we shall adopt another approach in order to determine whether the graceful exit from inflation takes place. In addition, the standard slow-roll breakdown argument does not hold true in the case of the cosmological evolution given in Eq.~(\ref{hub2}), since the first order slow-roll indices $\epsilon$ and $\eta$ never become of order one, as we now explicitly demonstrate. This, however, does not mean that inflation never ends, but it means that the slow-roll expansion does not provide enough information about graceful exit, at least in this case. In order to be sure, higher order terms should be calculated and then a more concise answer could be given. However, we will use another theoretical tool shortly to address the graceful exit issue. Before going into that, let us show that the slow-roll indices for the cosmology (\ref{hub2}), never actually become of order one, for the values of the parameters chosen as in Eq.~(\ref{parmchoice12}). In Fig.~\ref{plotslowroll}, we plotted the $N$-dependence of the first order slow-roll parameters $\epsilon$ and $\eta$, for the parameter values chosen as in Eq.~(\ref{parmchoice12}), and for $0\leq N \leq 60$. As we can see, both the first order slow-roll indices $\epsilon$ and $\eta$, take values well below unity for all the values of the $e$-foldings number $N$.
Moreover in Table \ref{table1} we present the values of the slow-roll parameters $\epsilon$ and $\eta$ for various values of the $e$-foldings number $N$. As it can be seen, the values of the slow-roll parameters are never close to one. Therefore, the graceful exit for the cosmology of Eq.~(\ref{hub2}) cannot be triggered by the breakdown of the slow-roll expansion, at least at leading order, with the leading order terms being actually the slow-roll parameters $\epsilon$ and $\eta$. 
\begin{table*}[h]
\small
\caption{\label{table1}Values of the slow-roll parameters $\epsilon$ and $\eta$ for various $N$ and for $\alpha=0.2$, $ \gamma=6.8$, $\beta=\frac{1}{42}$}
\begin{tabular}{@{}crrrrrrrrrrr@{}}
\tableline
\tableline
\tableline
Slow-roll Parameter & $N=0$  & $N=30$ & $N=50$ & $N=60$
\\\tableline
$\epsilon$ & $0.000727414$  & $0.0015348$ & $0.00257205$ & $0.00336111$
\\\tableline
$|\eta|$ & $0.0108638$  & $0.0096561$ & $0.00810471$ & $0.00692465$
\\\tableline
\tableline
 \end{tabular}
\end{table*}
Hence, our aim in this section is to demonstrate that graceful exit from inflation can actually occur for the inflationary cosmology of Eq.~(\ref{hub2}). Our theoretical argument is based on the approach adopted in Ref.~\cite{Bamba:2014jia}, in which case the exit from inflation is triggered by growing curvature perturbations, caused by $R^2$ correction terms. Particularly, as was shown in \cite{Bamba:2014jia} (see also \cite{Odintsov:2015cwa,Odintsov:2015wwp,Odintsov:2015ocy}), if the cosmological dynamical system reaches an unstable attractor, then the curvature perturbations grow at a sufficient rate to generate the graceful exit. Note that an unstable attractor in our case could be the unstable de Sitter attractor and also the appearance of the mimetic potential, instead of the $R^2$ correction terms, so our aim in this section is to show that there exist the unstable de Sitter vacua in the context of U-M gravity. Before getting into details, let us further support the theoretical argument of graceful exit from inflation via unstable final attractors. To this end, we make use of the Hamilton-Jacobi formulation of single field inflation models \cite{Lyth:2009zz}, since the resulting picture is qualitatively more or less the same. In the Hamilton-Jacobi formulation, the final attractor theorem states that all possible inflationary trajectories of the cosmological equations, quickly converge to a common attractor solution, if they are sufficiently close to each other initially. Quantitatively this means that if the final attractor solution, say $H_f$, is linearly perturbed, that is $H_f+\Delta H$, then the perturbation is decaying exponentially, in which case the final attractor solution is reached. In this case, the final attractor solution is stable towards linear perturbations. However, if the perturbations grow in time, then the final attractor solution is unstable, and this indicates that graceful exit is triggered for the inflationary solution. Practically, in single field inflation, this happens at the time instance that the slow-roll approximation breaks down and this is an indication of graceful exit. In the case at hand, the qualitative picture is the same, so we shall investigate if the de Sitter inflationary solutions of U-M gravity are unstable towards linear perturbations.  In section III we already demonstrated how the de Sitter cosmology can be realized in terms of the U-M gravity, so we use the results of this section. Therefore, we linearly perturb the differential equation (\ref{UM11}), by choosing the solution $H(t)$ to be of the form,
\begin{equation}\label{linpert}
H(t)=H_0+\Delta H(t)\, ,
\end{equation}
with $H_0$ the de Sitter solution we presented in section III. For the de Sitter case, the mimetic potential of the U-M gravity appears in Eq.~(\ref{mimeticpotdesitter}), so by substituting Eqs.~(\ref{mimeticpotdesitter}) and (\ref{linpert}) in the differential equation (\ref{UM11}), and by keeping linear terms and higher derivatives of the linear perturbation term $\Delta H(t)$, the differential equation (\ref{UM11}) becomes approximately,
\begin{equation}\label{diffeqnlinearpert}
\frac{36\,\,H_0^3 t}{\kappa ^2}\Delta H(t)-\frac{6 H_0}{\kappa ^2}\Delta \dot{H}(t)+\frac{4}{\kappa ^2}\Delta \ddot{H}(t)=0\, .
\end{equation}
Note that we kept leading order terms in the small $t$ limit, since we are interested in early times. 
\begin{figure}[h] \centering
\includegraphics[width=15pc]{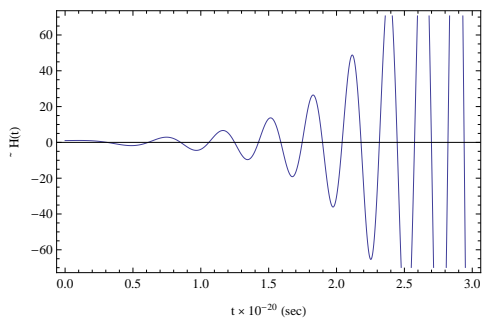}
\caption{Evolution of linear perturbations for the de Sitter solution of U-M gravity.}
\label{linpertev}
\end{figure}
The differential equation (\ref{diffeqnlinearpert}) can easily be solved and the solution is,
\begin{equation}\label{soldiffeqn}
\Delta H(t)=\e^{\frac{3 H_0 t}{4}} Ai\left(\frac{\frac{9 H_0^2}{16}-9 H_0^3 t}{3\ 3^{1/3} \left(H_0^3\right)^{2/3}}\right) C_1+\e^{\frac{3 H_0 t}{4}} Bi\left(\frac{\frac{9 H_0^2}{16}-9 H_0^3 t}{3\ 3^{1/3} \left(H_0^3\right)^{2/3}}\right) C_2\, ,
\end{equation}
with $C_1$, $C_2$ being arbitrary integration constants, while the functions $Ai(z)$ and $Bi(z)$ stand for the Airy functions. From the solution (\ref{soldiffeqn}) we can see that the perturbations grow in an oscillating way, which can also be seen in Fig.~\ref{linpertev}, where we plotted $\Delta H(t)$ as a function of the cosmic time $t$. Hence, since the linear perturbations of the de Sitter point $H_0$ of the form given in Eq.~(\ref{linpert}) grow in time, the de Sitter point is unstable and therefore if the de Sitter attractor is reached, graceful exit from inflation is triggered via the curvature perturbation mechanism of Ref.~\cite{Bamba:2014jia}. Notice that the cosmological evolution (\ref{hub2}), or in terms of the cosmic time (\ref{neweqns1}) can be brought into the de Sitter form, if $a_0\ll 1$ and the parameter $c_1$ is chosen to be $c_1=0$. Then, the scale factor of Eq.~(\ref{neweqns1}) becomes approximately $a(t)\sim \e^{H_0\,t}$, with $H_0=\beta\, \gamma$. Hence, our argument for the graceful exit holds true for the cosmological evolution (\ref{neweqns1}), or equivalently for (\ref{hub2}), which is also compatible with observations, as we evinced in the previous section.

\section{A Covariant Approach}

The presence of the unimodular constraint makes the unimodular mimetic gravity non-covariant, hence in this section we shall investigate how to construct a covariant unimodular mimetic gravity \cite{Buchmuller:1988yn,Henneaux:1989zc}. 
This can be achieved by the following covariant Lagrangian, 
\be
\label{MUR1}
S = \int d^4 x \left\{ \sqrt{-g} \left( \frac{R}{2\kappa^2}
 - \eta \left( \partial_\mu \phi \partial^\mu \phi + 1 \right) 
 - \lambda \right) + \lambda 
\epsilon^{\mu\nu\rho\sigma} \partial_\mu a_{\nu\rho\sigma} 
\right\}
+ S_\mathrm{matter} \left( g_{\mu\nu}, \Psi \right)\, ,
\ee
where $a_{\nu\rho\sigma}$ is a three-form field. The variation of the action (\ref{MUR1}) with respect to the three-form field $a_{\nu\rho\sigma}$ gives the equation
$0 = \partial_\mu \lambda$, which implies that $\lambda$ is a constant. On the other hand, the variation of the action (\ref{MUR1}) with respect to $\lambda$ gives,
\be
\label{LLUF19}
\sqrt{-g} = \epsilon^{\mu\nu\rho\sigma} \partial_\mu a_{\nu\rho\sigma}\, ,
\ee
which is actually a covariant version of the unimodular constraint (\ref{Uni1}). Owing to the fact that Eq.~(\ref{LLUF19}) can be solved with respect to $a_{\mu\nu\rho}$, there is no constraint on the metric $g_{\mu\nu}$. Then by varying the action (\ref{MUR1}), with respect to the metric tensor $g_{\mu\nu}$, we obtain,
\be
\label{MUR2}
0=\frac{1}{2}g_{\mu\nu} \left( R - 2\kappa^2 \lambda \right) - R_{\mu\nu} 
+ 2 \kappa^2 \eta \partial_\mu \phi \partial_\nu \phi
+ \frac{\kappa^2}{2} T_{\mu\nu} \, ,
\ee
where we have used the constraint given by the variation of the Lagrange multiplier $\eta$, 
\be
\label{MUR2x}
0=\partial_\mu \phi \partial^\mu \phi + 1\, .
\ee
By multiplying Eq.~(\ref{MUR2}) with $g^{\mu\nu}$ and using this constraint, we find 
an equation which determines $\eta$, which is, 
\be
\label{MUR2b}
2\kappa^2 \eta = R - 4\kappa^2 \lambda + \frac{\kappa^2}{2} T \, .
\ee
On the other hand, by varying again action(\ref{MUR1}) with respect to $\phi$, we obtain, 
\be
\label{MUR2c}
0 = \nabla^\mu \left( \eta \partial_\mu \phi \right)\, .
\ee
In the unimodular form of the FRW metric (\ref{UniFRW}), 
the non-zero components of the Ricci tensor are given in (\ref{Uni13}), 
while the Ricci scalar $R$ is given by (\ref{scalarunifrw}). 
We now assume that $\phi$ only depends on the time valuables, and in effect,  the constraint (\ref{MUR2x}) yields, 
\be
\label{MUR3}
\phi \left( \tau \right) = \int^\tau \frac{d\tau'}{a\left( \tau' \right)^3} \, .
\ee 
On the other hand, Because Eq.~(\ref{MUR2b}) tells that $\eta$ only depends on $\tau$, 
Eq.~(\ref{MUR2c}) takes the following form,
\be
\label{MUR4}
0 = \frac{d}{d\tau} \left( a \left(\tau\right)^3 \eta \right)\, ,
\ee
or equivalently, 
\be
\label{MUR5}
\eta = \eta_0 a\left( \tau \right)^{-3} \, ,
\ee
where $\eta_0$ is a constant. Then the $(t,t)$ and $(i,j)$ components of (\ref{MUR2}) have the following forms,
\begin{align}
\label{MUR6}
0 = & - 3 K^2 + \kappa^2 \lambda a\left( \tau \right)^{-6} 
+ 2\kappa^2 \eta_0 a\left( \tau \right)^{-9} + \kappa^2 \rho  a\left( \tau \right)^{-6} \, , \\
\label{MUR7}
0 = &  a\left( \tau \right)^{8} \left( 2 \dot K + 9 K^2 \right) 
 - \kappa^2 \lambda a\left( \tau \right)^2
+ \kappa^2 p a\left( \tau \right)^2\, ,
\end{align}
where $\rho$ is $p$ are the energy density and the pressure of matter fluids present. If we use the cosmological time $t$, because we have 
\be
\label{CUF6}
H = a^3 K\, , \quad \frac{dH}{dt} = a^6 \dot K + 3 a^6 K^2 \, , 
\ee
we can rewrite Eqs. (\ref{MUR6}) and (\ref{MUR7}) as follows, 
\begin{align}
\label{MUR8}
0 = & - 3 H^2 + \kappa^2 \lambda + 2\kappa^2 \eta_0 a\left( \tau \right)^{-3} 
+ \kappa^2 \rho  \, , \\
\label{MUR9}
0 = &  2 \dot H + 3 H^2 - \kappa^2 \lambda 
+ \kappa^2 p \, .
\end{align}
Eqs.~(\ref{MUR8}) and (\ref{MUR9}) are nothing but the standard FRW equations with 
the cosmological  constant $\kappa^2 \lambda$ and dark matter 
$2\kappa^2 \eta_0 a\left( \tau \right)^{-3}$.

Note that in order to realize an inflationary cosmology with graceful exit, as well as consistent with observational data, we need to add a scalar potential in the theory. Also we need to stress that the same FRW equations are obtained from our original non-covariant unimodular mimetic gravity formalism.

\section{Conclusions}

In this work we combined two conceptually different approaches of the Einstein-Hilbert gravity, the unimodular gravity
and mimetic gravity theories, in order to solve the cosmological constant problem and the dark matter problem in a unified geometrical way. By using the Lagrange multiplier method, we were able to materialize the unimodular and mimetic constraints at the Lagrangian level, and we used two Lagrange multipliers in order to achieve this. As we demonstrated, in the context of the unified theory of the two disciplines, which we called unimodular mimetic gravity, it is possible to realize a quite large number of cosmological evolutions, with some of them being exotic for the standard Einstein-Hilbert gravity. The equations of motion of the unimodular mimetic gravity theory constitute a reconstruction method, which when the Hubble rate is given, can be used to realize quite arbitrary cosmological scenarios. To this end we investigated how some well known cosmologies can be generated by unimodular mimetic gravity. Specifically, we realized the de Sitter cosmology, the Type IV singular cosmology, the $R^2$ inflation cosmology and also the cosmological evolution which corresponds to the standard Einstein-Hilbert cosmological evolution of a perfect fluid with constant equation of state. Also, by using the perfect fluid description, firstly introduced and employed in \cite{Bamba:2014wda}, we investigated how cosmologically viable cosmologies can be realized in the context of unimodular mimetic gravity. As we showed, cosmologies compatible with the latest Planck \cite{Ade:2015lrj,Planck:2013jfk} and even the BICEP2/Keck-Array data \cite{Array:2015xqh}, can be generated by the unimodular mimetic gravity framework. However, it is possible that in some cases, the graceful exit from inflation problem might exist. To this end, by using some qualitative arguments related to the final attractor theorem, we demonstrated that the graceful exit in the unimodular mimetic gravity can be triggered by curvature perturbations of the unstable de Sitter solutions. Specifically, as we showed, the unimodular mimetic theoretical framework leads to the unstable de Sitter solutions, which can generate the graceful exit from inflation. Finally, a covariant version of our theory was presented too.

A direct generalization of the unimodular mimetic gravity formalism, is to extend it in the context of modified gravity theories, for example the unimodular mimetic $F(R)$ gravity case, or even the unimodular mimetic $F(G)$ gravity case. In addition, it would be interesting to study solutions corresponding to compact objects, like relativistic stars or black holes, and check whether deviations from the predictions of general relativity exist. Also, in the unimodular case, the cosmological perturbations of unimodular gravity and the ordinary Einstein-Hilbert gravity are the same, at least when linear perturbations are considered, as was demonstrated in Refs.~\cite{Basak:2015swx,Gao:2014nia}. The question is whether this behavior persists in the case of the unimodular mimetic gravity. We hope to address some of these issues in detail in a future work.

In addition, a question that can be asked is why we did not use any matter fluids and we investigated the vacuum case of unimodular mimetic gravity? As we clearly explained in the text, the novel feature of our approach is that no matter fluids are required in order to realize various cosmological scenarios of the ordinary Einstein-Hilbert gravity, in which case their presence is compulsory. This is the new feature of our approach, that matter fluids are mimicked by the vacuum unimodular mimetic gravity, hence the geometry of the theory plays the role of the matter fluids. This feature of our theory, was also a feature of the mimetic gravity approach, and it is actually what justifies the terminology ``mimetic'', which means it mimics a cosmological behavior. The new feature of our theoretical framework is that we have more freedom in comparison to the standard mimetic approach. In principle, one could add a scalar field in the formalism, however, this could make things more complicated, since the presence of a scalar field would simply alter the functional form of the potential and the two Lagrange multipliers, in such a way so that the given cosmological evolution is realized.

\section*{Acknowledgments}

This work is supported by MINECO (Spain), project
  FIS2013-44881 and I-LINK 1019 (S.D.O),  by JSPS fellowship ID No.:S15127 (S.D.O.), and by Min. of Education and 
Science of Russia 
(S.D.O
and V.K.O) and  (in part) by
MEXT KAKENHI Grant-in-Aid for Scientific Research on Innovative Areas ``Cosmic
Acceleration''  (No. 15H05890) and the JSPS Grant-in-Aid for Scientific 
Research (C) \# 23540296 (S.N.).

\end{document}